\documentclass[11pt]{article}
\makeatletter
\usepackage[T1]{fontenc}
\usepackage[utf8]{inputenc}
\usepackage[english]{babel}
\usepackage{lmodern}
\usepackage{microtype}
\usepackage{csquotes}
\usepackage[a4paper,top=2.35cm,bottom=2.35cm,left=2.55cm,right=2.55cm,footskip=1.15cm]{geometry}
\usepackage{amsmath,amssymb,amsthm,mathtools}
\usepackage{booktabs,tabularx,longtable,array,multirow}
\newcolumntype{Y}{>{\raggedright\arraybackslash}X}
\newcolumntype{L}[1]{>{\raggedright\arraybackslash}p{#1}}
\usepackage{graphicx}
\usepackage{enumitem}
\usepackage{xcolor}
\definecolor{elpblack}{gray}{0.00}
\definecolor{elpgraydark}{gray}{0.25}
\definecolor{elpgraymid}{gray}{0.55}
\definecolor{elpgraylight}{gray}{0.96}
\usepackage[hyperfootnotes=false]{hyperref}
\usepackage[nameinlink,noabbrev]{cleveref}
\usepackage{titlesec}
\usepackage{fancyhdr}
\usepackage{setspace}
\usepackage{caption}
\usepackage{subcaption}
\usepackage{url}
\usepackage{float}
\usepackage{needspace}
\usepackage{tikz}
\usetikzlibrary{arrows.meta,positioning,calc,fit,backgrounds,decorations.pathreplacing,patterns}
\usepackage[
 backend=biber,style=authoryear,natbib=true,maxcitenames=2,maxbibnames=20,
 uniquename=false,doi=true,url=true,isbn=false,giveninits=true,dashed=false
]{biblatex}
\defbibheading{bibliography}[\refname]{\section*{#1}}
\hypersetup{hidelinks,pdfborder={0 0 0}}
\setlist{nosep,leftmargin=*}
\titleformat{\section}{\Needspace{7\baselineskip}\large\bfseries}{\thesection}{0.65em}{}
\titleformat{\subsection}{\Needspace{5\baselineskip}\normalsize\bfseries}{\thesubsection}{0.65em}{}
\titleformat{\subsubsection}{\normalsize\itshape}{\thesubsubsection}{0.65em}{}
\fancypagestyle{plain}{\fancyhf{}\fancyfoot[C]{\small\thepage}}
\newtheorem{theorem}{Theorem}[section]
\newtheorem{proposition}[theorem]{Proposition}

\theoremstyle{definition}
\newtheorem{definition}[theorem]{Definition}
\newtheorem{criterion}[theorem]{Criterion}

\theoremstyle{remark}

\newcommand{\paperversion}[1]{\gdef\PaperVersion{#1}}
\newcommand{\paperseries}[1]{\gdef\PaperSeries{#1}}
\newcommand{\papersubtitle}[1]{\gdef\PaperSubtitle{#1}}
\newcommand{\paperaffiliation}[1]{\gdef\PaperAffiliation{#1}}
\newcommand{\papercontact}[1]{\gdef\PaperContact{#1}}

\AtBeginDocument{%
  \providecommand{\PaperVersion}{draft}%
  \providecommand{\PaperSeries}{ForesightFlow Research \textperiodcentered{} Event-Linked Perpetuals}%
  \providecommand{\PaperSubtitle}{}%
  \providecommand{\PaperAffiliation}{}%
  \providecommand{\PaperContact}{}%
  \providecommand{\PaperStatus}{}%
}
\newcommand{\authoritynote}[1]{\begin{center}\begin{minipage}{0.92\linewidth}\small\textit{#1}\end{minipage}\end{center}}
\tikzset{
 pbox/.style={draw=black,rounded corners=2pt,align=center,inner sep=5pt,font=\footnotesize,text width=23mm,minimum height=9mm,fill=white},
 pwide/.style={draw=black,rounded corners=2pt,align=center,inner sep=5pt,font=\footnotesize,text width=31mm,minimum height=10mm,fill=white},
 psoft/.style={draw=black,dashed,rounded corners=2pt,align=center,inner sep=5pt,font=\footnotesize,text width=26mm,minimum height=9mm,fill=white},
 parr/.style={-{Latex[length=2.0mm]},line width=0.65pt,draw=black},
 pdash/.style={-{Latex[length=2.0mm]},line width=0.65pt,dashed,draw=black},
 pbrace/.style={decorate,decoration={brace,amplitude=4pt},line width=0.55pt,draw=black},
 ffbox/.style={pbox},ffsmall/.style={pbox,font=\scriptsize,inner sep=4pt},ffsoft/.style={psoft},ffarrow/.style={parr},ffdashed/.style={pdash}
}
\makeatletter
\renewcommand{\maketitle}{%
  \begin{center}
    {\small\scshape \PaperSeries\par}
    \vspace{0.8em}
    {\LARGE\bfseries \@title\par}
    \ifx\PaperSubtitle\@empty\else\vspace{0.35em}{\large \PaperSubtitle\par}\fi
    \vspace{0.75em}
    {\large \@author\par}
    \ifx\PaperAffiliation\@empty\else\vspace{0.18em}{\small \PaperAffiliation\par}\fi
    \ifx\PaperContact\@empty\else\vspace{0.12em}{\small \PaperContact\par}\fi
    \vspace{0.32em}{\small \@date\par}
  \end{center}
  \ifx\PaperStatus\@empty\else\authoritynote{\PaperStatus}\fi
  \vspace{0.25em}
}
\makeatother

\makeatother
\hypersetup{pdftitle={From Public Evidence to Contractual Outcome: First and Stable Decidability on Kalshi},pdfauthor={Maksym Nechepurenko},pdfsubject={Event-Linked Perpetuals --- Kalshi Research Track},pdfkeywords={prediction markets, contractual decidability, adjudication, public evidence, Kalshi}}
\paperseries{ForesightFlow Research \textperiodcentered{} Event-Linked Perpetuals \textperiodcentered{} Paper 7.3}
\paperversion{r0.9.3}
\papersubtitle{First and Stable Decidability on Kalshi}
\title{From Public Evidence to Contractual Outcome}
\author{Maksym Nechepurenko}
\paperaffiliation{Research Department of Devnull FZCO, Dubai, UAE}
\papercontact{maksym@devnull.ae \quad ORCID 0000-0002-9515-8841}
\date{September 2026 \textbar{} Version r0.9.3}
\begin{document}
\maketitle
\begingroup
\renewcommand{\thefootnote}{}
\footnotetext{\footnotesize\textit{ELP --- Kalshi Research Track.} This paper forms part of the Kalshi-focused research track within the Event-Linked Perpetuals series.}
\endgroup
\begingroup\footnotesize
\begin{abstract}
Public evidence can become sufficient to settle a prediction-market contract before the venue records its first determination, but the relevant boundary depends on the applicable rule version, exact release object, source hierarchy, correction history, and unfinished contract conditions. This paper defines two Kalshi clocks: \emph{first decidability}, the earliest contemporaneous singleton in the rule--evidence mapping, and \emph{stable decidability}, the retrospective earliest time after which the same singleton remains unchanged through finalization.

A completed retrospective identification test establishes a narrow but consequential feasibility result. In a frozen blind pilot, all 25 identities and blinding checks passed and current rule text was recovered for all 25; no exact or bounded historical rule version and no exact or bounded official source-release object was recovered. The full historical recovery covered 152,694 ordinary tickers, 11,530 exact event identities, and 6,540 read-only official requests, with zero historically eligible events and zero historically eligible tickers. This is an observability result, not a claim that no market was decidable or that public evidence never existed.

A completed prospective infrastructure shakedown established observation capability for three source programmes across 25 markets, with integrity revalidation of 781,266 lifecycle frames, 22 closed lower-bounded reconnect receipts, no unresolved reconnect gap, no due-but-missed official release, and an inactive price layer. Production evidence enrollment is active. The prospective sample is constructed only at enrollment close from prospectively frozen identities and pre-outcome fields; its frozen target size is selected mechanically under the registered full, reduced, exploratory, or no-go support disposition. No contractual-decidability clock, human-adjudication, price, or cross-venue result is reported here.
\end{abstract}
\endgroup
\noindent\textbf{Keywords:} prediction markets; contractual decidability; adjudication; public evidence; Kalshi.\\
\textbf{JEL:} G13, G14, G18.

\section{Introduction}
A real-world event, a source publication, a contractually sufficient evidence set, and a venue determination are different objects. They may occur close together, but no identification strategy may assume that they do. Close-to-determination time can include source delay, contractual waiting, ambiguity, venue review, or unobserved procedure. A useful decomposition requires a clock upstream of venue action.

The phrase ``the outcome was knowable'' is too loose. This paper instead studies public contractual sufficiency under the applicable recorded rules. A market is first-decidable when contemporaneously available admissible evidence reduces the contract's settlement mapping to one value and no contract-required future condition remains. Because source corrections or rule clarifications can reopen that mapping, the paper separately defines a retrospective stable clock. This distinction prevents an early but later reversed singleton from being treated as permanent certainty.

The problem is related to semantic non-fungibility: apparently similar event claims can differ in source, temporal scope, exceptions, and resolution semantics \parencite{gebele2026semantic}. It is also distinct from settlement manipulation, which concerns incentives to alter the evidence-generating process rather than when the archived evidence becomes sufficient under the contract \parencite{dai2026settlement}.

Manual adjudication is costly, so the paper does not promise exchange-wide clocks. It constructs a reproducible probability sample with explicit weights, blind review, and a measured attrition ladder. An inability to identify a clock is a property of public governance observability, not a missing value to be replaced by the venue's own timestamp.

Settlement legibility has recently been used to explain which uncertainties become prediction-market contracts at all \parencite{adegbenro2026formation}. This paper moves to a later and more granular boundary: conditional on a listed contract and its contemporaneous rule graph, when does the public evidence set first and stably imply one admissible settlement value? The first/stable distinction and blind upstream adjudication are the paper's core conceptual claims.

\section{Rule--evidence model}
For market $i$, let $\Gamma_i(t)$ be the rule graph applicable at valid time $t$, $\mathcal E_i(t)$ the admissible public evidence objects available by $t$, and $\mathcal Y_i$ the settlement-value set. Define
\[
\Psi_i(t)=\Psi_i\bigl(\Gamma_i(t),\mathcal E_i(t)\bigr)\subseteq\mathcal Y_i.
\]
The graph includes contract/series terms, market-specific primary and secondary rules, exact source declarations, threshold and time-zone predicates, early-close clauses, and dated additional details. Precedence is encoded only when the recorded materials establish it.

\begin{definition}[First decidability]
\[
t^{first}_{decide,i}=\inf\{t:|\Psi_i(t)|=1\ \text{and no contract-required future condition remains at }t\}.
\]
The definition is contemporaneous. It does not assume that a later source correction is impossible.
\end{definition}

\begin{definition}[Stable decidability]
Let $y_i^*$ be the final archived settlement value. Stable decidability is the retrospective clock
\[
t^{stable}_{decide,i}=\inf\{t:\Psi_i(s)=\{y_i^*\}\ \text{for every observed rule/source version }s\in[t,t_{final,i}]\}.
\]
It is a robustness clock, not a real-time information set available to traders.
\end{definition}

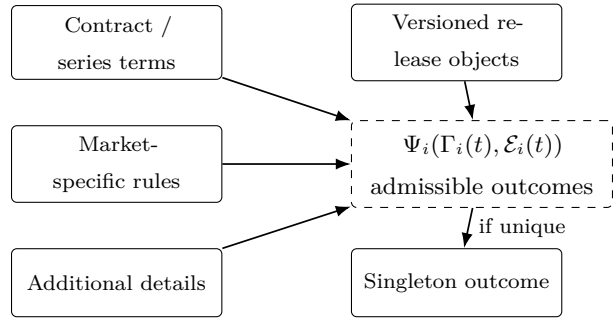
\begin{figure}[H]
\centering
\begin{tikzpicture}[node distance=6mm and 17mm]
\node[ffsmall,text width=25mm](terms){Contract / series terms};
\node[ffsmall,text width=25mm,below=of terms](market){Market-specific rules};
\node[ffsmall,text width=25mm,below=of market](details){Additional details};
\node[ffsmall,text width=25mm,right=of terms](source){Versioned release objects};
\node[ffsoft,text width=31mm,right=of market](psi){$\Psi_i(\Gamma_i(t),\mathcal E_i(t))$\\admissible outcomes};
\node[ffsmall,text width=25mm,right=of details](single){Singleton outcome};
\foreach \u in {terms,market,details,source}{\draw[ffarrow](\u)--(psi);}
\draw[ffarrow](psi)--node[right,font=\scriptsize]{if unique}(single);
\end{tikzpicture}
\caption{Contractual-decidability graph. A named source is one input; the rule--evidence mapping determines whether the admissible outcome set is a singleton.}
\label{fig:73-rule}
\end{figure}

\begin{proposition}[Named-source insufficiency]
A source name and publication timestamp do not generally identify contractual decidability.
\end{proposition}
Different strikes, exception clauses, time windows, source hierarchies, or pending future conditions can map the same release to different admissible outcome sets.

\begin{proposition}[Interval evidence induces latency bounds]
If $t_{decide}\in[a_i,b_i]$ and $t_{det}$ is exact, then
\[
t_{det}-t_{decide}\in[t_{det}-b_i,\ t_{det}-a_i].
\]
No point estimate is identified without an additional model for the location of the clock inside the interval.
\end{proposition}

\begin{criterion}[No determination-defined decidability]
Venue determination or final outcome may not be used to construct the blind upstream clock. Doing so would mechanically force determination latency toward zero and introduce hindsight.
\end{criterion}

\section{Retrospective identification test}
The historical ordinary-market frame remains a venue-side feasibility denominator. It contains 152,694 exact ordinary tickers and 11,530 exact event identities; the completed recovery issued 6,540 read-only official requests. In the frozen 25-market blind pilot, 25/25 identities were auditable, 25/25 blinding checks passed, and current rule text was recovered for 25/25. Yet exact or bounded historical rule versions were recovered for 0/25 and exact or bounded official source-release objects for 0/25. Across the full historical recovery, the number of historically eligible events is zero and the number of historically eligible tickers is zero.

The registered public historical record therefore did not identify the version-bound rule object and official release object required to reconstruct the contractual-evidence clock at the accepted evidence grade. This finding does not say that the contracts were not decidable, and it does not assert that public evidence never existed. It says that the required historical binding cannot be reconstructed from the available public record. The historical pilot is complete and is not an active 500-market execution plan.

\section{Prospective evidence-capture amendment}
The prospective infrastructure shakedown started at $T_{C0}=\text{2026-09-01T22:31:38Z}$ and passed all gates at $T_{\mathrm{PROD}}=\text{2026-09-04T22:54:26Z}$ after 260,568 seconds (72 hours 22 minutes 48 seconds) of valid observation. It established observation capability for prospectively frozen market, event, and series identities; English-locale rule versioning; official source-release observation; exact and bounded publication-time semantics; and separate rule/source, venue-outcome, price, and concurrent public metadata layers. The shakedown validated infrastructure and adapter capability to preserve source corrections and revisions. It covered the BLS CPI, FOMC target-rate, and Federal Register executive-order programmes across three events and 25 markets. These programmes demonstrate observation capability only and are not an empirical cohort; the Federal Register subset was conformance-only.

The shakedown recorded 781,266 lifecycle frames, completed a full integrity revalidation, classified 22 reconnect receipts as closed lower bounds, and found no unresolved reconnect gap. There were no due-but-missed official releases, and the price layer remained empty. Concurrent public Polymarket metadata produced 1,405 preliminary candidates, none semantically accepted. Sustainability checks passed for 30, 42, and 60 days. These facts establish observation capability, not a source-programme sample or a contractual-decidability result.

Prospective enrollment begins at $T_{\mathrm{PROD}}$, 2026-09-04T22:54:26Z. The primary enrollment interval starts at that instant and ends immediately before 2026-10-04T22:54:26Z. The registered 42-day total enrollment ceiling is 2026-10-16T22:54:26Z, a possible 12-day extension beyond the 30-day primary enrollment rather than an additional 42-day extension. The event horizon is 2026-10-19T22:54:26Z; hard follow-up ends 2026-11-03T22:54:26Z. Shakedown observations are excluded by default unless a separately frozen no-outcome-leakage rule passes. Only post-enrollment receipts may support the prospective study.

\section{Sample design and prospective frame}
The final prospective frame is built at enrollment close. It uses prospectively frozen identities and only pre-outcome source class, market form, and expected horizon. Venue outcome, determination, finalization, endpoint class, and price path are excluded from sampling and initial blind adjudication.

Before outcome inspection, the registered support disposition selects a frozen target $n^*$: full support sets $n^*=500$ when at least 500 blind-adjudicable markets, 100 independent release/event clusters, ten rule/series families, three source classes, and 80 percent exact-or-bounded rule/source support are present; reduced support sets $n^*=150$ when at least 150 markets and 40 clusters are present; exploratory support sets $n^*=50$ when at least 50 markets and 20 clusters are present; otherwise $n^*=0$ and the disposition is no-go. In a simple random full-support sample, $n^*=500$ gives a worst-case 95 percent margin of error of approximately $0.044$ for a proportion; stratification, clustering, weights, and adjudication attrition can widen that precision. These rules are frozen before outcome inspection.

\subsection{Pre-outcome strata and deterministic allocation}
The primary stratum is constructed from variables known without venue outcome or post-outcome latency: source-class bucket, binary/scalar market form, and expected horizon from the recorded pre-outcome schedule. Sparse cells are deterministically merged into an explicit residual cell within market form before allocation. Let $N_h$ be the resulting stratum size. Allocation uses square-root size weights,
\[
 a_h=\sqrt{N_h},\qquad n_h^{(0)}=n^*\frac{a_h}{\sum_k a_k},
\]
with a minimum of two observations for an eligible non-empty stratum, a cap at $N_h$, and deterministic largest-remainder adjustment until $\sum_h n_h=n^*$. The inclusion probability is $\pi_i=n_h/N_h$ and the analysis weight is $1/\pi_i$.

Within each stratum, prospectively enrolled market identities are ordered by
\[
\operatorname{DeterministicOrder}(s_{prod},\,\text{market identifier}),
\]
where the deterministic production seed is defined in the sampling protocol supplied with the replication materials. The lowest $n_h$ values are selected. The algorithm is specified here; the actual registry is generated at enrollment close and receives an inclusion-probability audit.

\subsection{Pilot and independent second review}
A completed 25-market historical feasibility pilot tested packet construction, rule-version recovery, exact release-object recovery, packet blinding, time per packet, clock grades, and disagreement coding; it is not merged into the prospective frame. A future prospective process check, if required, is governed by the same no-outcome-leakage and independent-selection rules.

At least 20 percent of the sample, with a target minimum of 100 eligible packets, receives independent second review. Operationally, the independently selected count is $n_{\mathrm{second}}=\min\{n^*,\max(\lceil0.20n^*\rceil,\min(100,n^*))\}$, so it cannot exceed the realised sample and 100 remains a target rather than an achieved result. The subset is selected by deterministic order using a third seed defined in the sampling protocol. A separate conflict-enriched audit can be added, but it is not mixed with the probability-based reliability subset.

\section{Population and lifecycle authorities for sampling}
The KMVE V1 authority closes the seven-day MVE market-object population at 7,611,594 exact REST tickers, 5,262,526 exact event keys, three current-validated collection identities, and 83,701 selected-leg primitives \parencite{nechepurenko2026kmve}. These objects are useful for dependence diagnostics and a small exploratory MVE stratum, not for an unrestricted child-ticker adjudication sample.

The historical venue-side frame provides a feasibility denominator but does not construct contractual-decidability clocks or a historically admissible clock cohort. Paper~7.1 supplies the 152,694 ordinary-market frame and lifecycle/endpoint authority, while Paper~7.2 supplies the MVE hierarchy relevant to dependence and pseudoreplication \parencite{elp71preprint,elp72preprint}. The prospective programme provides infrastructure capability and an active enrollment design; it does not yet report clock values. Table~\ref{tab:73-readiness} separates completed evidence from the registered follow-up.

\begin{table}[H]
\centering\footnotesize
\caption{Evidence availability for Paper 7.3 after the historical feasibility test and prospective shakedown.}
\begin{tabularx}{\textwidth}{@{}L{0.27\textwidth}L{0.22\textwidth}Y@{}}
\toprule
Layer & Current state & Consequence \\
\midrule
Historical feasibility denominator & 152,694 exact ordinary tickers; 11,530 exact event identities & Completed recovery; not a prospective sampling frame \\
Historical rule/source binding & 0 historically eligible events; 0 historically eligible tickers & No historical contractual-decidability cohort at the accepted evidence grade \\
MVE population hierarchy & 7,611,594 objects with exact parent/leg fields & Exploratory event/collection stratum; no unrestricted child-ticker sampling \\
Prospective shakedown & 72-hour capability check passed across three source programmes & Supports enrollment, not a clock estimate \\
Prospective enrolled frame & Built at enrollment close if support thresholds pass & Frozen $n^*$ sample or registered reduced, exploratory, or no-go disposition \\
First/stable decidability & Not yet adjudicated prospectively & Core Paper 7.3 outcome remains to be measured \\
Blind packet and review protocol & Specified & Human adjudication begins only after registered prospective frame verification \\
\bottomrule
\end{tabularx}
\label{tab:73-readiness}
\end{table}

\subsection{Blinding boundary}
Sampling and initial blind review may use exact identity, recorded pre-outcome rule fields, source declarations, expected horizon, and market form. They may not expose the venue result, determination time, finalization time, endpoint class, determination-to-endpoint duration, or later price path. After the initial review record is preserved, later linkage reveals venue clocks and final value for consistency checks and latency construction. A contradiction creates a new adjudication version; it never overwrites the initial record.

\subsection{No broad rules crawl}
The packet builder uses prospectively archived versioned rule objects, prospectively archived official source objects, contemporaneous receipts, and revision/supersession graphs for the selected sample. Later retrieval may verify current availability; it cannot create missing valid-time evidence after the fact. The 7.6-million-object MVE population is not a rules-crawl denominator.

\section{Blind adjudication protocol}
Adjudication proceeds in two phases.

\textbf{Initial blind review.} The reviewer receives exact identities, rule versions, candidate source declarations, and archived source releases, but not Kalshi's result, determination timestamp, or finalization timestamp. The reviewer selects the applicable release object, evaluates each rule predicate, and records first and stable decidability as an exact time, interval, proxy, unmeasured, or conflict.

\textbf{Later venue linkage.} After the initial review record is preserved, the venue clocks and final value are introduced. Contradictions create a new adjudication version; they do not overwrite the initial review.

At least 20 percent of the sample, with a target minimum of 100 eligible packets, receives independent second review. The operational count is bounded by the realised sample, so the target minimum is not described as achieved when fewer than 100 eligible packets are available. A conflict-enriched secondary audit is reported separately and is not mixed with the random reliability subset. Agreement is reported as raw agreement and Cohen's $\kappa$ for nominal fields \parencite{cohen1960agreement}; clock agreement uses interval overlap and absolute boundary differences. Prevalence-sensitive interpretation of $\kappa$ is explicit.

Each packet records source-object identity, publication time/interval, rule predicates, candidate outcome set, first/stable clocks, exclusion category, reviewer, and version lineage.

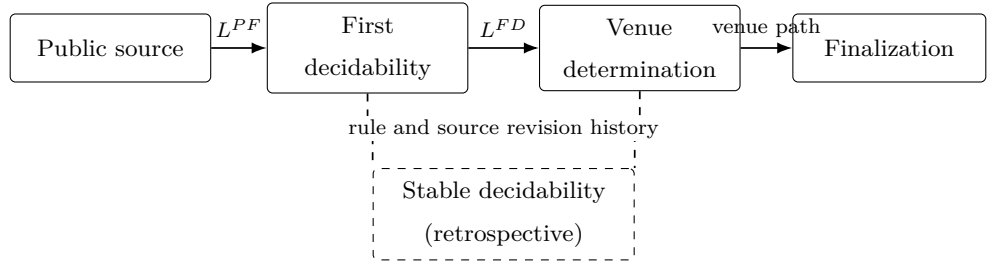
\begin{figure}[H]
\centering
\begin{tikzpicture}[x=1cm,y=1cm]
\node[pbox,text width=23mm] (src) at (-5.2,1.1) {Public source};
\node[pbox,text width=23mm] (first) at (-1.8,1.1) {First\\decidability};
\node[pbox,text width=23mm] (det) at (1.8,1.1) {Venue\\determination};
\node[pbox,text width=22mm] (fin) at (5.1,1.1) {Finalization};
\draw[parr] (src.east)--node[above,font=\scriptsize]{$L^{PF}$}(first.west);
\draw[parr] (first.east)--node[above,font=\scriptsize]{$L^{FD}$}(det.west);
\draw[parr] (det.east)--node[above,font=\scriptsize]{venue path}(fin.west);
\node[psoft,text width=31mm] (stable) at (0,-1.1) {Stable decidability\\(retrospective)};
\draw[dashed,line width=.65pt] (first.south) -- ++(0,-0.42) -| (stable.north west);
\draw[dashed,line width=.65pt] (det.south) -- ++(0,-0.42) -| (stable.north east);
\node[font=\scriptsize,align=center,fill=white,inner sep=1pt] at (0,0.03) {rule and source revision history};
\node[font=\scriptsize,align=center,text width=105mm] at (0,-2.6) {Stable decidability is retrospective; its timestamp may precede or follow venue determination.};
\end{tikzpicture}
\caption{First and stable decidability under a partial order. Source publication precedes first decidability under the accepted packet, and venue determination precedes finalization. Stable decidability can fall before or after venue determination; the dashed relations are analytical dependencies, not chronological arrows.}
\label{fig:73-clocks}
\end{figure}

\section{Estimands and partial identification}
Primary latencies are
\[
L_i^{FD}=t_{det,i}-t^{first}_{decide,i},\qquad
L_i^{SD}=t_{det,i}-t^{stable}_{decide,i},
\]
with publication-to-first-decidability
\[
L_i^{PF}=t^{first}_{decide,i}-t_{pub,i}
\]
and stability gap
\[
V_i=t^{stable}_{decide,i}-t^{first}_{decide,i}.
\]
If stable decidability occurs after venue determination, $L^{SD}<0$ is not interpreted as negative delay; it is a conflict class indicating that the venue acted before the publicly archived rule--evidence sequence became retrospectively stable.

For $t_{close}<t_{det}$ and $t^{first}_{decide}\le t_{det}$, the bounded share of close-to-determination time occurring after first decidability is
\[
Q_i^{post}=\frac{t_{det,i}-\max(t_{close,i},t^{first}_{decide,i})}{t_{det,i}-t_{close,i}}\in[0,1].
\]
This replaces an unbounded ratio that could exceed one when decidability preceded close.

Exact and interval clocks are reported separately. For threshold $x$, interval observations yield lower and upper indicators
\[
I_i^L(x)=1[\overline L_i\le x],\qquad I_i^U(x)=1[\underline L_i\le x],
\]
which bound the share determined within $x$ units of decidability. Midpoints are absent from primary analysis.

The attrition ladder is
\[
N_{sample}\rightarrow N_{named\ source}\rightarrow N_{release\ object}\rightarrow N_{publication}\rightarrow N_{first}\rightarrow N_{stable}\rightarrow N_{latency}.
\]
It is reported by stratum with sampling weights. Cluster bootstrap uses source release as the primary dependence unit, with event/series sensitivity.

\section{Registered empirical questions}
\textbf{Q1 --- Identification yield.} What weighted share of sampled markets produces an exact first clock, interval first clock, stable clock, or no admissible clock? No adjective such as ``nontrivial'' is specified in advance.

\textbf{Q2 --- Positive residual latency.} What is the identified share for which $L^{FD}$ exceeds the timestamp/interval precision tolerance? Exact and bounded shares are reported.

\textbf{Q3 --- Revision sensitivity.} How often is $V_i>0$, and which source or rule mechanisms account for the gap between first and stable decidability?

\textbf{Q4 --- Heterogeneity.} How do identification yield and $L^{FD}$ vary by source class, market form, horizon, and pre-outcome rule features? These are descriptive, with multiplicity correction or hierarchical shrinkage if many contrasts are retained.

\textbf{Q5 --- Reviewer reproducibility.} What are raw agreement, $\kappa$, clock-interval overlap, and adjudicated boundary differences on the random audit subset?

Pseudo-source releases and generic homepage timestamps provide negative controls. Venue close is an anchor substitution, while determination-as-decidability is shown only as a mechanical pathology, not a competing estimator.

\section{Validity and reproducibility}
The design requires reproducible sampling without post-outcome timing fields, recorded rule versions, exact source-object identity or an explicit failure classification, blind initial adjudication, and independent review. A market lacking the required timing evidence is excluded from the corresponding timing analysis while remaining visible in the observability and attrition accounting.

The analysis distinguishes sample and weights, source-object attrition, first/stable clock grades, reviewer reliability, latency bounds, source/rule heterogeneity, revisions and conflict cases, and negative controls.

\section{Limitations}
Contractual sufficiency can be genuinely ambiguous. Public source pages can be revised or overwritten, and the venue may use evidence not visible to the researcher. Stable decidability is retrospective and must not be described as traders' real-time knowledge. Human adjudication remains judgmental despite blinding and replication. A stratified sample supports sample-weighted claims, not automatic exchange-wide generalization when coverage or nonresponse is uneven. Long $L^{FD}$ is a public-evidence residual, not a causal estimate of venue inefficiency.

\section{Conclusion}
The completed historical test establishes that a clock may be economically well-defined yet retrospectively unobservable from the available public historical record. The historical ordinary frame remains an important venue-side feasibility denominator, but it cannot supply the version-bound rule object and official release object required for the registered contractual-evidence clock. The KMVE hierarchy remains available for a separately clustered exploratory MVE extension without allowing millions of related child objects to dominate the statistical unit.

The prospective shakedown demonstrates source-observation capability and active evidence enrollment. What remains is the paper's core semantic measurement task: blind reviewers must reconstruct exact rule versions and public release objects, identify first and stable decidability, and preserve interval, unmeasured, and conflicting cases. The final prospective sample is conditional on registered support thresholds at enrollment close. No human adjudication, clock value, latency estimate, price study, or accepted cross-venue pair is reported.

This separation is the identification strategy. Population scale, venue finality, and public contractual sufficiency are distinct layers. No endpoint count, determination timestamp, or final result can substitute for a reproducible rule--evidence proof.

\appendix
\section{Adjudication classifications}
The minimum taxonomy is grouped as follows:
\begin{itemize}
\item exact or interval first/stable clocks;
\item source found with rule ambiguity, and rule found with source unavailable;
\item multiple plausible releases, conflicting primary evidence, and unmeasured cases.
\end{itemize}
Reconciliation creates a new version and never deletes initial reviewer judgments.

\section*{Data and Code Availability}
The registered MVE population and hierarchy are available as the \emph{Kalshi Multivariate Event Market Materialization Dataset (KMVE)}, Mendeley Data, V1, doi: \href{https://doi.org/10.17632/fn65786cg6.1}{10.17632/fn65786cg6.1}. The manuscript describes the prospective sampling and adjudication design. Raw active production-enrollment capture is not distributed, and no empirical production-enrollment result dataset yet exists. No sampled ticker identities, contractual-decidability clocks, adjudication outcomes, price results, or accepted cross-venue pairs are released with this preprint.

\section*{Generative AI Disclosure}
OpenAI ChatGPT and Codex were used for editorial and technical assistance during manuscript preparation. The author made all substantive research decisions, reviewed the final manuscript, and assumes full responsibility for its contents.

\section*{Funding}
This research received no external funding.

\section*{Competing Interests}
The author is affiliated with the Research Department of Devnull FZCO and leads the ForesightFlow research programme. No external sponsor influenced the research design, analysis, interpretation, or decision to publish. The article does not evaluate a commercial product or make investment recommendations.

\printbibliography[heading=bibliography,title={References}]
\end{document}